\documentclass{article}
\usepackage{graphicx}
\graphicspath{%
    {converted_graphics/}
    {C:/Dropbox/1-kgl-top/Papers/2014/CorrespondenceVaporLiquid/}
}
\begin{document}

\begin{center}
{\LARGE Toward a Comprehensive Model of Snow Crystal Growth:}\vskip6pt

{\LARGE \ 3. The Correspondence Between Ice Growth from Water Vapor}\vskip6pt

{\LARGE and Ice Growth from Liquid Water}\vskip10pt

{\Large Kenneth G. Libbrecht}\vskip4pt

{\large Department of Physics, California Institute of Technology}\vskip-1pt

{\large Pasadena, California 91125}\vskip-1pt

\vskip18pt

\hrule\vskip1pt \hrule\vskip14pt
\end{center}

\textbf{Abstract.} We examine ice crystal growth from water vapor at
temperatures near the melting point, when surface premelting creates a
quasiliquid layer at the solid/vapor interface. Recent ice growth
measurements as a function of vapor supersaturation have demonstrated a
substantial nucleation barrier on the basal surface at these temperatures,
from which a molecular step energy can be extracted using classical
nucleation theory. Additional ice growth measurements from liquid water as a
function of supercooling exhibit a similar nucleation barrier on the basal
surface, yielding about the same molecular step energy. These data suggest
that ice growth from water vapor and from liquid water are both well
described by essentially the same underlying nucleation phenomenon over a
substantial temperature range. A physical picture is emerging in which
molecular step energies at the solid/liquid, solid/quasiliquid, and
solid/vapor interfaces create nucleation barriers that dominate the growth
behavior of ice over a broad range of conditions. Since the step energy is
an equilibrium quantity, just as surface melting is an equilibrium
phenomenon, there exists a considerable opportunity to use many-body
simulations of the ice surface structure and energetics at equilibrium to
better understand many dynamical aspects of ice crystal growth.

\section{Introduction}

Recent measurements of ice growth rates from water vapor have shown a
characteristic nucleation-limited behavior, with the growth velocity normal
to the surface exhibiting the exponential dependence $v_{n}\sim \exp
(-\sigma _{0}/\sigma ),$ where $\sigma $ is the water vapor supersaturation
relative to ice at the surface and $\sigma _{0}$ is a parameter extracted
from the data \cite{kglalphas13}. Classical nucleation theory allowed the
authors to calculate the free energy (per unit length) of a molecular step
on the crystal surface from the measured $\sigma _{0}(T),$ thus yielding the
step energy as a function of temperature for the basal and prism facets.
Near the melting point, specifically at $T=-2$ C, values of $\sigma
_{0}\approx 0.4$ percent on the basal surface and $\sigma _{0}<0.01$ percent
on the prism surface were reported. The substantially larger nucleation
barrier on the basal surface provides a simple explanation for the formation
of thin plate-like crystals at this temperature, an observation first made
nearly 75 years ago \cite{nakaya54}.

It has similarly been long known that growing free-standing ice crystals in
liquid water at small undercoolings results in the formation of thin
plate-like crystals \cite{hobbs74}, and quantitative measurements of the
basal growth velocity over a half-century ago yielded a similar exponential
dependence $v_{n}\sim \exp (-\Delta T_{0}/\Delta T),$ where $\Delta
T=T_{m}-T_{surf}$ is the supercooling at the growing interface, $T_{m}$ is
the melting temperature, and $\Delta T_{0}\approx 0.24$ C is an
experimentally determined constant \cite{Hillig58}. This functional form is
again indicative of a nucleation barrier, and the molecular step energy can
be calculated from $\Delta T_{0}.$ Similar measurements for the growth of
the prism surface yielded a much lower nucleation barrier, again consistent
with the formation of thin plate-like crystals.

It is also well known that surface premelting is an important structural
feature of the ice solid/vapor interface \cite{dash06}. In essence, the
molecular layers near the crystalline surface are not as tightly bound as in
the bulk, resulting in the formation of an amorphous \textquotedblleft
quasiliquid\textquotedblright\ layer (QLL) at the interface. The detailed
properties of the quasiliquid layer are not well known, but it is generally
believed that the layer thickness is strongly temperature dependent,
diverging to infinite thickness as the melting point is approached, and that
the quasiliquid properties become similar to those of bulk water as the
layer thickness increases. If true, then we would expect that the
solid/quasiliquid interface should resemble the solid/liquid interface as
the melting point is approached.

It is significant, therefore, that the measurements of $\sigma _{0}$ and $%
\Delta T_{0}$ cited above yield essentially the same molecular step energy,
suggesting that the dynamical growth behaviors in both these cases arise
from essentially the same physical phenomenon. To our knowledge, this
correspondence between ice growth from water vapor and from liquid water has
not yet been examined in the scientific literature. The data indicate that
the solid/liquid and solid/quasiliquid interfaces yield comparable step
energies over a substantial temperature range, a somewhat predictable
statement that nevertheless was apparently not anticipated in previous
investigations of surface premelting or ice crystal growth.

Below we examine these points in detail, using basic thermodynamics to
connect ice growth from water vapor and ice growth from liquid water. We
further suggest that many-body molecular-dynamics simulations of the \textit{%
equilibrium} structure and energetics of the ice surface, including step
energies and premelting, could be used to better understand the \textit{%
non-equilibrium} dynamics of ice crystal growth over a broad range of
conditions.

\section{Nucleation Theory}

For a simple monomolecular solid surface in contact with its pure melt or
pure vapor, crystal growth is driven by a chemical potential jump at the
interface -- by a nonzero supercooling at the solid/liquid interface or by a
nonzero supersaturation in the case of growth from vapor. For both these
cases, a classical polynuclear growth model gives the normal growth velocity 
\cite{saito96} 
\begin{equation}
v_{n}\approx A_{0}\Delta \mu \exp \left( \frac{-S\beta ^{2}a^{2}}{\Delta \mu
kT}\right) 
\end{equation}%
where $S\approx 1$ is a geometrical factor, $\beta $ is the step free energy
(per unit length) at the crystal interface, $a$ is the molecular size, $%
\Delta \mu $ is the chemical potential jump at the interface, $k$ is the
Boltzmann factor and $T$ is temperature. The prefactor $A_{0}$ is somewhat
model dependent, as the many-body microphysics at the interface is complex
and not well understood, but it is expected that $A_{0}$ has at most a weak
dependence on $\Delta \mu .$ We chose the functional form above so that $%
v_{n}\sim \Delta \mu $ (the Wilson-Frenkel law \cite{saito96}) when $\Delta
\mu $ is large enough that the exponential term goes to unity, as this is a
generally accepted description of crystal growth in the absence of a
nucleation barrier.

For ice growth from water vapor, we have $\Delta \mu \approx \sigma kT$ for $%
\sigma \ll 1,$ where $\sigma =(c_{surf}-c_{sat})/c_{sat}$ is the
supersaturation with respect to ice just above the growing surface, $%
c_{surf} $ is the water vapor number density at the surface, and $c_{sat}(T)$
is the equilibrium number density above a flat ice surface \cite{saito96}.
For nucleation-limited growth we write%
\begin{eqnarray}
v_{n} &\approx &A_{vap}\sigma e^{-\sigma _{0}/\sigma } \\
\sigma _{0} &=&\frac{S\beta _{vap}^{2}a^{2}}{k^{2}T_{m}^{2}}  \nonumber
\end{eqnarray}%
where $\beta _{vap}$ is the step energy at the solid/vapor interface. The
prefactor can be obtained from ideal-gas statistical mechanics when there is
no kinetic barrier (Hertz-Knudsen growth \cite{saito96}), giving 
\begin{equation}
A_{vap}=v_{kin}=\frac{c_{sat}}{c_{solid}}\sqrt{\frac{kT}{2\pi m}}
\end{equation}%
where $c_{solid}=\rho _{ice}/m$ is the number density for ice. Note that the
growth rate is often written as $v_{n}=\alpha v_{kin}\sigma $, where $\alpha
\leq 1$ is known as the attachment coefficient \cite{libbrechtreview05}.

For growth from liquid, $\Delta \mu \approx \ell t$ for $t\ll 1$, where $%
t=(T_{m}-T)/T_{m}$ is the dimensionless supercooling and $\ell $ is the
solid/liquid latent heat per molecule $(\ell =1.0\times 10^{-20}$ J for
ice), giving 
\begin{eqnarray}
v_{n} &\approx &A_{liq}te^{-t_{0}/t}  \label{liquidgrowtheq} \\
t_{0} &=&\frac{S\beta _{liq}^{2}a^{2}}{\ell kT_{m}}  \nonumber
\end{eqnarray}%
The prefactor is determined by how fast liquid molecules diffuse into
position to join the solid lattice, which can be estimated \cite{saito96}%
\begin{equation}
A_{liq}\approx \frac{\ell }{6\pi a^{2}\eta _{eff}}
\end{equation}%
where $\eta _{eff}$ is the effective viscosity for liquid near the surface.
The kinetics of liquid water near an ice surface is nontrivial, and it is
possible that $\eta _{eff}$ may differ from the normal bulk viscosity near
the melting point, $\eta _{0}\approx 1.8\times 10^{-3}$ Pa-s. Nevertheless,
assuming $\eta _{eff}\approx \eta _{0}$ and using $a\approx 0.32$ nm gives $%
A_{liq}\approx 3$ m/sec. An alternative model based on the density
functional theory of freezing yields $A_{liq}\approx (kT_{m}/m)^{1/2}\approx
350$ m/sec \cite{chernov91}.

In the case of ice growth from water vapor in the presence of surface
premelting, we assume that any nucleation barrier must exist at the
solid/QLL interface, and at some point the quasiliquid layer would be thick
enough that $\beta _{vap}\approx \beta _{liq}.$ In this case we obtain 
\begin{eqnarray}
\sigma _{0} &\approx &\frac{\ell }{kT_{m}}t_{0}  \label{sig0} \\
&\approx &2.6t_{0}  \nonumber
\end{eqnarray}%
or equivalently $\sigma _{0}\approx 0.01\Delta T_{0},$ where $\Delta
T_{0}=t_{0}T_{m}$ is expressed in degrees C.

\begin{figure}[ht] 
  \centering
  \includegraphics[width=3.0in,keepaspectratio]{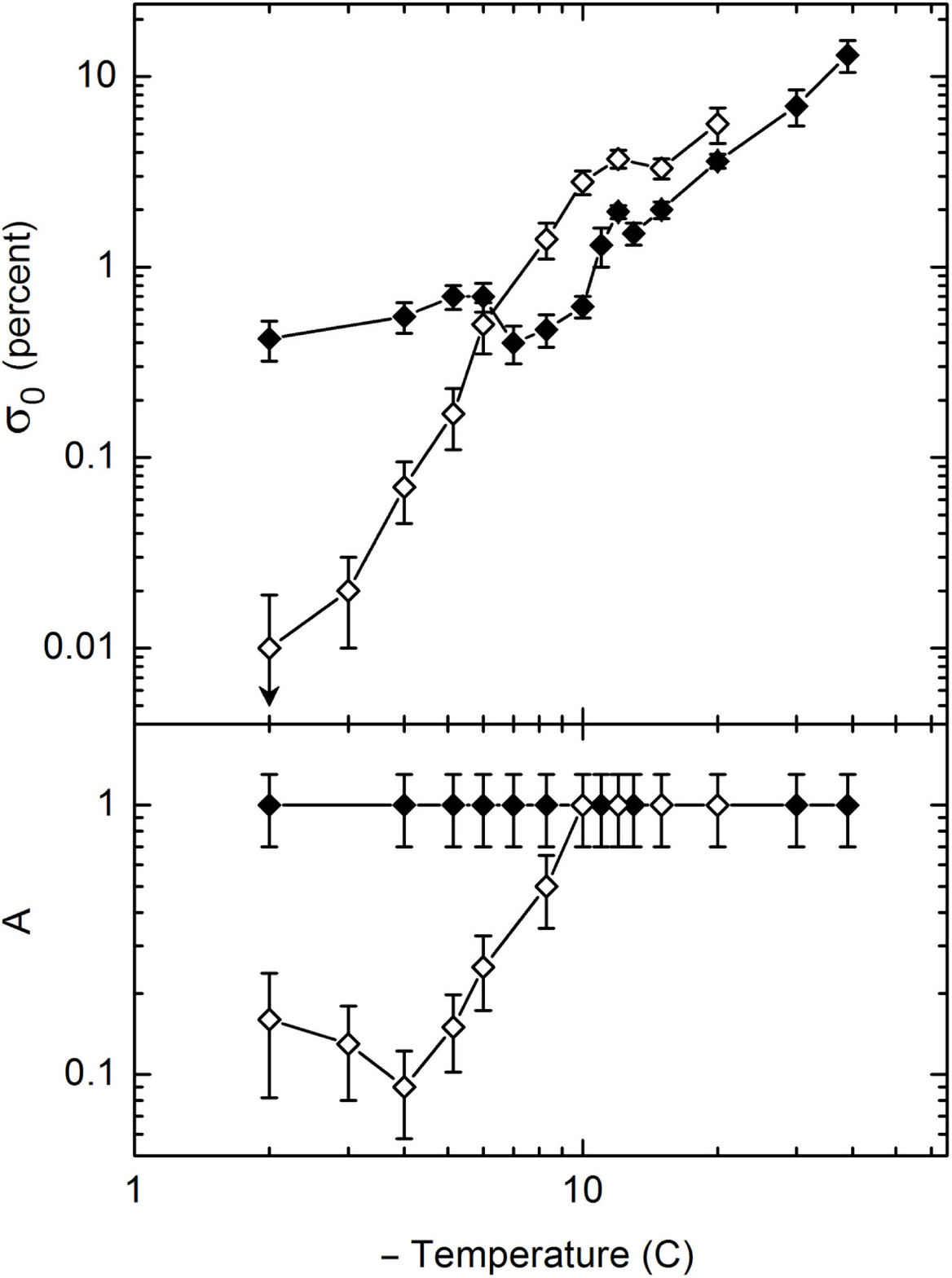}
  \caption{Measurements of the intrinsic
growth rates of the principal ice crystal facets from water vapor. The
growth velocity normal to the surface is described by $v=\protect\alpha %
v_{kin}\protect\sigma _{surf}$, where $\protect\sigma _{surf}$ is the
supersaturation at the surface and the attachment coefficient is
parameterized with $\protect\alpha (T,\protect\sigma _{surf})=A\exp (-%
\protect\sigma _{0}/\protect\sigma _{surf}).$ The solid points show the
measured $A(T)$ and $\protect\sigma _{0}(T)$ for the basal facets, while the
open points describe measurements of the prism facets, from \protect\cite%
{kglalphas13}.}
  \label{intrinsic}
\end{figure}

\section{Experimental Data}

\subsection{Growth from Water Vapor}

The experimental data from \cite{kglalphas13} are the most recent, most
extensive, and we believe the most accurate measurements to date of the
intrinsic ice growth rates of faceted basal and prism surfaces from water
vapor. In particular, the experiments were performed at low pressures to
reduce the effects of particle diffusion through the surrounding gas, and
special attention was paid to determining $\sigma _{surf},$ the
supersaturation at the growing surface. Over the temperature range $-2$ C $%
\geq T\geq -40$ C, the data are well described by a dislocation-free
nucleation-limited crystal growth model, parameterized using $v_{n}=\alpha
v_{kin}\sigma _{surf}$ with $\alpha (\sigma _{surf},T)=A\exp (-\sigma
_{0}/\sigma _{surf})$. The measured parameters $A(T)$ and $\sigma _{0}(T)$
for the basal and prism facets are shown in Figure \ref{intrinsic}.

Our understanding of these data is quite crude, in part because our overall
understanding of the many subtleties of crystal growth dynamics is somewhat
rudimentary, especially in a material like ice that exhibits substantial
surface premelting. In \cite{comprehensive12} we attempted to construct a
comprehensive physical picture of ice growth from water vapor, connecting
the growth measurements shown in Figure \ref{intrinsic} with related
morphological observations. Since our present purpose is to examine the
correspondence between growth from water vapor and from liquid water, we
will focus on the growth from vapor at $T=-2$ C, which is the highest
temperature for which we have data in Figure \ref{intrinsic}.

\begin{figure}[t] 
  \centering
  \includegraphics[width=2.5in, keepaspectratio]{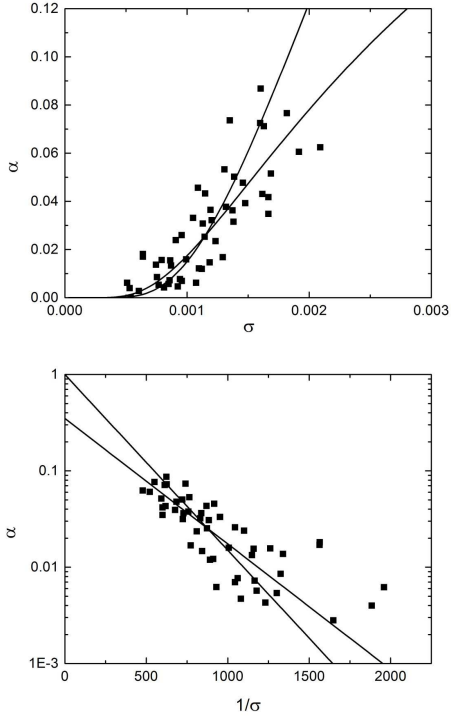}
  \caption{Measurements of the
condensation coefficient $\protect\alpha $ for ice growth from water vapor,
for the basal facet at -2 C as a function of near-surface supersaturation $%
\protect\sigma $. The same data points are displayed two ways in the above
plots. Both plots use the parameterizations $v_{n}=\protect\alpha v_{kin}%
\protect\sigma $ and $\protect\alpha =A\exp (-\protect\sigma _{0}/\protect%
\sigma ),$ from \protect\cite{kglalphas13}. The two fit lines show $\protect%
\alpha =\exp (-0.0042/\protect\sigma )$ and $\protect\alpha =0.35\exp
(-0.003/\protect\sigma ).$ The lowest-$\protect\sigma $ points were given a
lower weight in the fits, as the growth was slowest for these points and we
believed most susceptible to systematic errors. The two fit lines both
describe the data reasonably well, roughly indicating the degree of
uncertainty in extracting the parameters $A$ and $\protect\sigma _{0}$ from
the ice growth data.}
  \label{vapordata}
\end{figure}

We focus especially on the basal facet growth at $T=-2$ C, and these data
are shown in more detail in Figure \ref{vapordata}. As was described in \cite%
{kglalphas13}, the prefactor $A$ at this temperature was in part
extrapolated from lower temperatures, since the basal data were consistent
with $A=1$ over the entire temperature range of the measurements. This
assumption of $A=1$ gave the fit parameters $(A,\sigma _{0})=(1,0.0042)$
shown in Figure \ref{intrinsic}. As is shown in Figure \ref{vapordata},
however, the data at $T=-2$ C are also reasonably well fit using $(A,\sigma
_{0})=(0.35,0.003),$ and this figure demonstrates the uncertainty in
extracting these parameters from the growth measurements. The data clearly
indicate a nucleation-limited growth behavior, and from the fits we obtain $%
\sigma _{0}=0.42\pm 0.15$ percent, the uncertainty including our best
estimate of possible systematic effects. The experiment and data analysis
are described in more detail in \cite{kglalphas13}.

\subsection{Growth from Liquid Water}

Although numerous observations of ice growth from liquid water have been
reported in the literature, in most cases the growth rates are strongly
limited by the diffusion of latent heat generated at the growing crystal
surface, and in this case the bath supercooling $\Delta T_{\infty }$ can be
markedly different from $\Delta T_{surf}$ at the interface. Since the
interfacial temperature is the essential variable governing the growth
dynamics, we restrict our attention to experiments where heat-diffusion
effects have been carefully considered and modeled to determine $%
v_{n}(\Delta T_{surf}).$

\begin{figure}[tbp] 
  \centering
  \includegraphics[width=3.83in,height=3.01in,keepaspectratio]{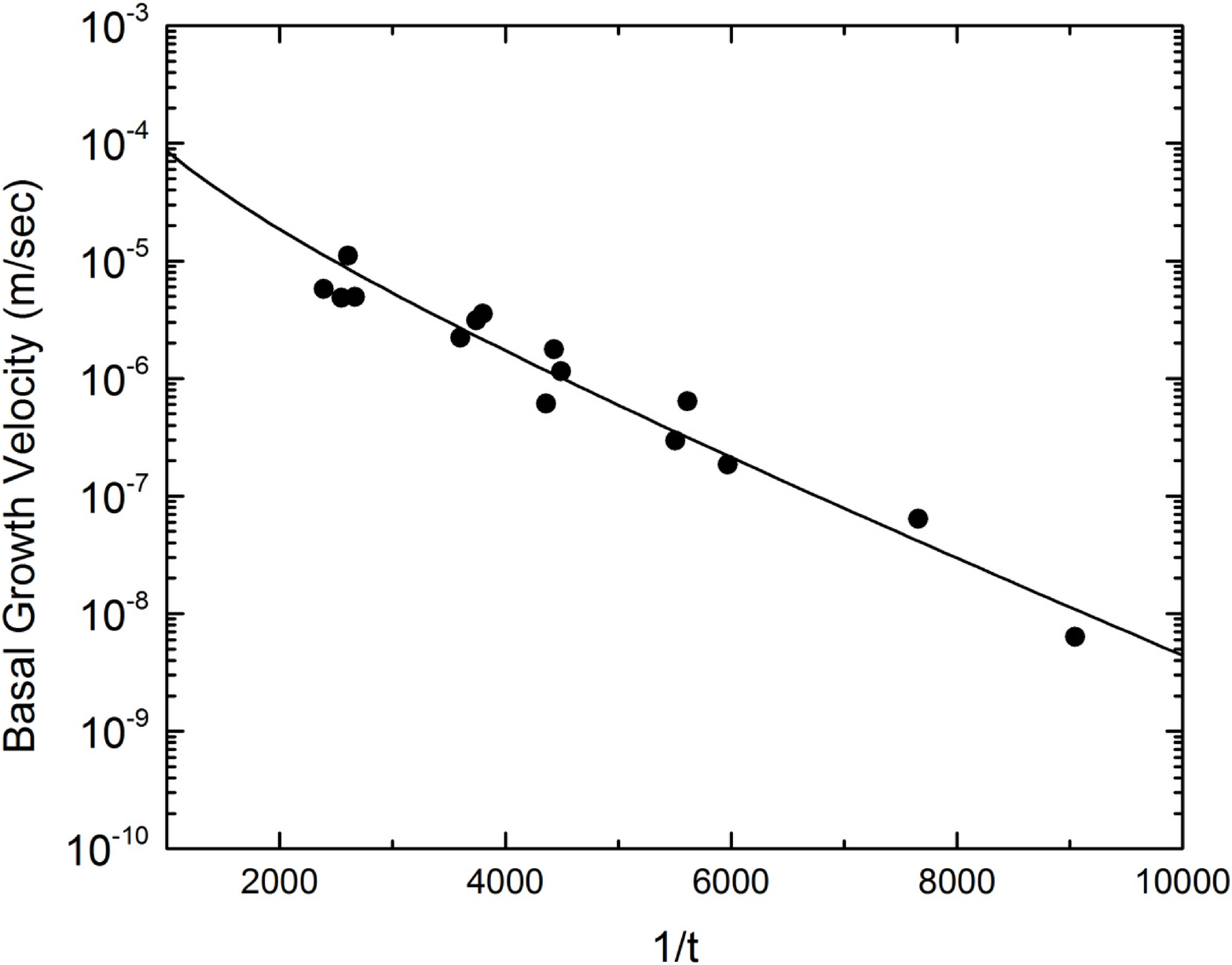}
  \caption{Measurements of the basal growth
velocity as a function of $t^{-1},$ where $t$ is the dimensionless
supercooling, for ice growth from liquid water. The data are reproduced from 
\protect\cite{Hillig58}, and the fit line is $v_{n}=0.2t\exp (-0.00084/t),$
giving $\Delta T_{0}=tT_{m}=0.24$ C.}
  \label{liquidgrowth}
\end{figure}

The work presented in \cite{Hillig58} is the best we have found describing
measurements of the basal growth rate as a function of supercooling, and the
relevant data from this source are reproduced in Figure \ref{liquidgrowth}.
In this series of measurements, the ice growth was observed in a thin
capillary tube, allowing $\Delta T_{surf}$ to be determined with good
accuracy, and repeated zone refining was used to remove impurities from the
water sample (it was found that impurities reduced the measured growth
rates).

From the liquid growth data shown in Figure \ref{liquidgrowth}, together
with the assumed functional form given in Equation \ref{liquidgrowtheq}, we
extract the value $\sigma _{0}=\ell t_{0}/kT_{m}=0.24\pm 0.04$ percent.
Comparing with $\sigma _{0}=0.42\pm 0.15$ percent from the vapor growth
data, we see that the two values are consistent with being the same to
within the estimated uncertainties. This supports our hypothesis that ice
growth from water vapor and ice growth from liquid water are related near
the melting point, when significant surface premelting is present. In
particular, these data show that the step free energy at the solid/QLL
interface at $T=-2$ C is quite close to the step energy at the solid/liquid
interface near $T=0$ C. Extrapolating the basal $\sigma _{0}$ data in Figure %
\ref{intrinsic} to higher temperatures, we expect $\sigma _{0}$ to approach $%
\sigma _{0}\approx 0.24\pm 0.04$ percent as the temperature approaches the
melting point.

The fit value $A_{liq}\approx 0.2$ m/sec from Figure \ref{liquidgrowth} is
low compared to our estimate above using the viscosity of bulk liquid water,
but this may be unique to the basal surface. Similar measurements described
in \cite{Hillig56} found growth rates as high as $v_{n}=0.077$ m/sec at a
supercooling of $\Delta T=10$ C for nonbasal surfaces having a much lower
nucleation barrier. Since thermal diffusion and any residual nucleation
barrier both reduce the growth rate, this $v_{n}$ measurement suggests a
lower limit of $A_{liq}>2$ m/sec for nonbasal surfaces. Additional
measurements in \cite{shibkov05} describe ice dendrite tip velocities as
high as 0.6 m/sec at $\Delta T_{\infty }=30$ C, again giving a lower limit
of $A_{liq}>5$ m/sec for nonbasal surfaces at high growth rates.

On the prism surface, both ice growth from liquid water and ice growth from
water vapor show much smaller nucleation barriers. The measured $\sigma _{0}$
at $-2$ C (see Figure \ref{intrinsic}) and $t_{0}$ near the melting point 
\cite{Hillig56} are both essentially indistinguishable from zero. The story
is not as compelling as for the basal case, but nevertheless the data are
consistent with the correspondence we are proposing. Both the basal and
prism data suggest a smooth transition from solid/liquid growth to
solid/vapor growth near the melting point. At lower temperatures the QLL
becomes thinner and eventually disappears entirely, resulting in the complex
growth behavior described in more detail in \cite{comprehensive12}.

\section{Discussion}

The data described above suggest a rather simple picture of ice crystal
growth from water vapor in the presence of surface premelting. At low
supersaturations on the basal surface, the growth rate is strongly limited
by the nucleation of molecular terraces at the solid/quasiliquid interface.
Near the melting point, the step energy at this interface is approximately
equal to the step energy at the solid/liquid interface. Thus there is a
simple correspondence between ice growth from water vapor and ice growth
from liquid water.

On the prism surface, or on the basal surface at high supersaturations, the
growth is not strongly limited by a nucleation barrier. In this case the
growth kinetics at the solid/quasiliquid interface are much faster than at
the quasiliquid/vapor interface, as indicated by the fact that $A_{liq}\gg
A_{vap},$ or equivalently $A_{liq}\gg v_{kin},$ by a large factor. In this
situation the growth is limited by kinetics at the quasiliquid/vapor
interface.

Assuming an essentially liquid surface, we therefore would expect $A\approx 1
$ for ice growth from water vapor \cite{watercondcoef86}. The data in Figure %
\ref{intrinsic} are consistent with this except on the prism facet at high
temperatures. We have no explanation for this discrepancy. The physical
picture described above is quite compelling, however, enough so that it
suggests that there may be some currently unrecognized systematic error in
the measurements. Additional data or analysis will likely be necessary to
resolve this issue.

The data also suggest rather clearly that the key to understanding ice
crystal growth is understanding the step energies on the different surfaces
as a function of temperature. To my knowledge, relatively little attention
has been given to using many-body molecular dynamics simulations to examine
molecular step energies, although simulations of premelting and ice growth
have been studied by a number of researchers \cite{kusalik12, MD09, pan11,
conde08, furukawa97, moldymice05}. 

The fact that the step energy is an equilibrium quantity means that its
calculation from molecular dynamics simulations could be substantially
simpler than simulating full growth dynamics, as only equilibrium energetics
calculations are needed. However, the measured step energies are quite low
compared to expectations for a sharp molecular step ($\beta \ll \beta _{0}$
in \cite{kglalphas13}). A simple geometrical argument then suggests that
surface relaxation \textquotedblleft smooths out\textquotedblright\ the
terrace edge over perpendicular distances of perhaps $20a$ near the melting
point to minimize the overall surface energy in the vicinity of the step. If
true, then simulations must include large surfaces to fully model the step
energy. This smoothing is substantially less at lower temperatures, however,
making the calculations relatively easier. 

Regardless of the details, it appears there exists a substantial opportunity
to use current many-body molecular dynamics simulations, with perhaps only
minor modifications, to calculate step energies at the ice surface. If
successful, these calculations could go far toward furthering our
understanding of the dynamics of ice crystal growth. 

\bibliography{kglbiblio3}
\bibliographystyle{unsrt}

\end{document}